\documentclass[fleqn,usenatbib]{mnras}

\usepackage{newtxtext,newtxmath}

\usepackage[T1]{fontenc}
\usepackage{natbib}
\usepackage{wrapfig}
\usepackage{longtable}
\usepackage{enumitem}
\usepackage{multirow}
\usepackage{subfig}
\usepackage[toc,page]{appendix}
\usepackage{graphicx}
\usepackage{epstopdf}
\usepackage{pdfpages}
\usepackage{float}
\usepackage{upgreek}
\DeclareRobustCommand{\VAN}[3]{#2}
\let\VANthebibliography\thebibliography
\def\thebibliography{\DeclareRobustCommand{\VAN}[3]{##3}\VANthebibliography}

\newcommand{\microns}{\,$\upmu\mathrm{m}$ } 

\title[Super-resolving {\it Herschel} imaging]{Super-resolving {\it Herschel} imaging: a proof of concept using Deep Neural Networks}

\author[L. Lauritsen et al.]{
Lynge Lauritsen$^{1}$\thanks{E-mail: lynge.lauritsen@open.ac.uk},
Hugh Dickinson$^{1}$,
Jane Bromley$^{2}$,
Stephen Serjeant$^{1}$, 
Chen-Fatt Lim$^{3,4}$,
\vspace*{0.25cm}\\ 
{\LARGE\rm Zhen-Kai Gao$^{4,5}$, Wei-Hao Wang$^{4}$}
\\
$^{1}$School of Physical Sciences, Faculty of Science, Technology, Engineering \& Mathematics, The Open University, \\
Walton Hall, Kents Hill, Milton Keynes, MK7 6AA, United Kingdom\\
$^{2}$School of Computing \& Communications, Faculty of Science, Technology, Engineering \& Mathematics, The Open University, \\
Walton Hall, Kents Hill, Milton Keynes, MK7 6AA, United Kingdom\\
$^{3}$Graduate Institute of Astrophysics, National Taiwan University, Taipei 10617, Taiwan\\
$^{4}$Academia Sinica Institute of Astronomy and Astrophysics (ASIAA), No. 1, Section 4, Roosevelt Road, Taipei 10617, Taiwan\\
$^{5}$Graduate Institute of Astronomy, National Central University, Taoyuan 32001, Taiwan\\}

\date{Accepted XXX. Received YYY; in original form ZZZ}

\pubyear{2021}

\begin{document}
\label{firstpage}
\pagerange{\pageref{firstpage}--\pageref{lastpage}}
\maketitle

\newcommand{\bfreferee}[0]{}

\begin{abstract}
Wide-field sub-millimetre surveys have driven many major advances in galaxy evolution in the past decade, but without extensive follow-up observations the coarse angular resolution of these surveys limits the science exploitation.
This has driven the development of various analytical deconvolution methods. In the last half a decade Generative Adversarial Networks have been used to attempt deconvolutions on optical data. Here we present an autoencoder with a novel loss function to overcome this problem in the sub-millimeter wavelength range. This approach is successfully demonstrated on {\it Herschel} SPIRE {\bfreferee 500\microns} COSMOS data, with the super-resolving target being the JCMT SCUBA-2 {\bfreferee 450\microns} observations of the same field. We reproduce the JCMT SCUBA-2 images with high fidelity using this autoencoder. {\bfreferee This is quantified through} the point source {\bfreferee fluxes and positions, the completeness and the purity.}  
\end{abstract}




\section{Introduction}\label{Introduction}

All astronomical imaging has an intrinsic angular resolution limit, whether due to seeing, diffraction, instrumental effects or (in the case of an interferometer) the longest available baselines. In single-dish diffraction-limited imaging, there is formally no signal on Fourier scales smaller than the diffraction limit. {\bfreferee This is because Fraunhofer diffraction is mathematically equivalent\footnote{The Fourier transform of the aperture gives the amplitude pattern of a point source, e.g. a 1D top-hat aperture yields a sinc function. Incident energy is proportional to amplitude squared, e.g. a top-hat aperture yields sinc$^2$. For a 2D circular aperture this is  $\mathrm{sinc}^2(|{\bf r}|)$, i.e. an Airy function.} to a Fourier transform, so the large-scale boundary of the telescope aperture also implies there is no image information smaller than some angular scale.} {\bfreferee For} interferometers, the Fourier plane has incomplete coverage especially approaching the smallest angular scales.

{\bfreferee However}, there are often strong scientific drivers for improving angular resolution. {\bfreferee Among many advantages, higher angular resolution affords the possibility of more reliable multi-wavelength cross-identifications \citep[e.g.][]{Franco2018,Dudz2020}, improved deblending of nearby sources \citep[e.g.][]{Hodge2013,Simpson2015}, and fainter fundamental confusion limits. For example, \citet[]{SCUBA-2_legacy_2013} used the better angular resolution of the James Clerk Maxwell Telescope (JCMT) SCUBA-2 $450$\microns data compared to the {\it Herschel} SPIRE instrument to probe sub-millimetre (sub-mm) number counts with fluxes below 20\,mJy where source confusion becomes problematic in {\it Herschel} SPIRE data \citep{Herschel_Number_Count_2010, Herschel_DR1_2016}. Further, \citet[]{SCUBA-2_legacy_2013} also resolved a larger part of the Cosmic Infrared Background than that possible using {\it Herschel} SPIRE. There has therefore} been a great deal of interest in developing algorithms for recovering or estimating some of the missing {\bfreferee Fourier} data on smaller {\bfreferee angular} scales (see \cite{Deconvolution_review_2002} and references therein for a detailed discussion), including approaches that exploit abundant multi-wavelength data where that exists {\bfreferee \citep[e.g.][]{Hurley2017,Jin2018}}. 

One domain where angular resolution gains are particularly advantageous is {\bfreferee sub-mm} astronomy. Wide-field extragalactic surveys have proved transformative for e.g. nearby galaxies \citep[e.g.][]{Dustpedia}, galaxy evolution \citep[e.g.][]{Lutz2014,SMG_Hierachy_Universe,Geach2017MNRAS.465.1789G} and strong gravitational lensing \citep[e.g.][]{Negrello2010}. 
Sub-mm galaxies can also be used to trace possible protoclusters through 
overdensities \citep{cluster_core, indicate, protocluster}. Much of this progress has been driven by surveys with the SPIRE instrument \citep{Griffin2010} on the ESA {\it Herschel}\footnote{Herschel is an ESA space observatory with science instruments provided by European-led Principal Investigator consortia and with important participation from NASA.} mission \citep{Pilbratt2010}, but at moderately high redshifts (e.g. $z\stackrel{>}{_\sim}4$) the detections tend to be dominated by the longest wavelength band ($500$\microns) where the diffraction-limited point spread function (PSF) has a full width half maximum (FWHM) of $36.6''$. Higher resolution mapping is possible with ground-based facilities such as \mbox{SCUBA-2} \citep{Holland2013} and the Atacama Large Millimeter Array (ALMA), but the mapping efficiencies are far lower and it is not feasible to map the entire {\it Herschel} SPIRE extragalactic survey fields with sub-mm ground-based facilities to comparable depths. Furthermore, the abundant multi-wavelength data available for multi-wavelength prior based deconvolution work in the deeper {\it Herschel} fields \citep[e.g.][]{Oliver2012} does not exist at equivalent depths for all wider-area {\it Herschel} surveys \citep[e.g.][]{Eales2010}.

In the past half-decade the use of machine learning, and in particular Convolutional Neural Networks (CNNs), has gained popularity as a potential solution to
image deconvolution
\citep{GalGAN, image_restoration_cycle, Moriwaki_2021}. These CNNs all use a Generative Adversarial Neural Network (GAN) for their CNN based image restoration. A GAN consists of two neural networks called the generator, and the discriminator. The generator is trained to generate an image that looks {\bfreferee ``}realistic{\bfreferee'' (according to some relevant quantitative metric)}, while the discriminator will try to determine if a given image is real or generated \citep{goodfellow2014generative}. As they are trained {\bfreferee concurrently with competing objectives} the performance of the generator will depend on the detailed characteristics of the discriminator. This paper presents an alternative approach using an autoencoder\footnote{The architecture of an {\bfreferee autoencoder} is similar to that of many GANs \cite{GalGAN, image_restoration_cycle, Moriwaki_2021} but does not use a discriminator network output as part of its objective or loss functions.} with a specially designed loss function. Similar networks have been used to enhance and remove noise from astronomical images at other wavelengths \citep[e.g.][]{2020MNRAS.tmp.3376V}.

Architecturally, an autoencoder contains an encoding CNN which extracts a relatively small number of scalar-valued features from input images. The values of these features are referred to collectively as an \textit{embedding} of the input image. A second, decoding CNN is then used to generate an image with the same dimensions as the input images, using only the information encoded by the embedding. The objective of the decoding network is to produce an image that closely matches a given target image that is associated with the corresponding input. During training, the encoding network learns to construct an embedding which optimally represents the features of the input image that are required for the decoding network to generate a close match to the corresponding target \citep{Goodfellow-et-al-2016}. This paper will show that a simple {\bfreferee autoencoder} network can be used to super-resolve {\it Herschel} SPIRE data, and achieve angular resolution comparable to that of JCMT. This super-resolved data can then be used to determine the sky locations and fluxes of previously 
{\bfreferee lower-resolution observations of} 
sub-mm galaxies. 

In \S\ref{Training Data} two different training sets are discussed, in \S\ref{Network Architecture} the network architecture, and the loss function is described, in \S\ref{Results} the network performance on both observed, and simulated data is presented, and finally \S\ref{Discussion}, and \S\ref{Conclusion} will discuss and summarise the network performance. 

\section[training Data]{Training Data}\label{Training Data}

The autoencoder presented in this paper is a \textit{supervised} machine learning algorithm. Supervised learning requires the use of a training dataset with known truth values. Two separate training sets were used to train the network presented in this paper: (i) a simulated training set{\bfreferee,} made using images generated by a modified version of the Empirical Galaxy Generator (EGG) software \citep{EGG} as both the the target and input images, and (ii) using the JCMT SCUBA-2 $450$\microns maps from the STUDIES project \citep{STUDIES} as target examples, {\bfreferee with} the {\it Herschel} SPIRE maps for the COSMOS field as input images \citep{Levenson2010,Oliver2012,Viero2013}. Table \ref{tab:telescope_instruments} shows the FWHM, confusion limit and pre-interpolation pixel scales of the instruments {\bfreferee whose data are} used in this paper.

\begin{table*}
\centering
\caption{\label{tab:telescope_instruments}
Characteristics of the Herschel SPIRE and JCMT SCUBA-2 instruments. Confusion limits and PSF widths are from
\citet{Herschel_confusion}, 
\citet{STUDIES} and \citet{SCUBA-2_2}.
}
\begin{tabular}{|l||l|l||l||l|}
\hline
	 Characteristic & \multicolumn{3}{c}{Herschel SPIRE} & JCMT
	 SCUBA-2\\
	 \hline
	 Wavelength & $250$\microns & $350$\microns & $500$\microns & $450$\microns \\
	 \hline
	 PSF FWHM & 18.1"& 24.9" & 36.6" &7.9"\\
	 \hline
	 Confusion noise ($\sigma$, mJy/beam) & $5.8\pm 0.3$ & $6.3\pm 0.4$ & $6.8\pm 0.4$ & $1$\\
	 \hline
	 Pixel scale & $6$" & $8.33$" & $12$" & $1$" \\

\hline

\end{tabular}
\end{table*} 

\subsection[Simulated Training]{Simulated Data}\label{Simulated Data}

There are very few large astronomical fields that have been surveyed by both {\it Herschel} SPIRE and JCMT SCUBA-2. Accordingly, simulations must be used to create a large representative dataset of images to train the network. The simulated dataset was generated using a version of The Empirical Galaxy Generator (EGG) software {\bfreferee (see \cite{EGG} for an in-depth discussion on the workings of EGG)}, that was modified to {\bfreferee avoid simulating galaxies with negligible} 
infrared fluxes {\bfreferee using an empirically determined bolometric luminosity threshold imposed within the code}. This was done to improve efficiency and objects with negligible FIR luminosity were not simulated. This modification {\bfreferee altered the number counts outside the FIR range, but  reproduced realistic number counts in the FIR range at a lower computational cost. We verified that it} made no discernible difference to the output images, while saving considerable computation time. EGG is designed to generate a mock survey catalog with realistic multi-wavelength galaxy number counts, {\bfreferee using an empirical calibration}, and with realistic galaxy clustering. {\bfreferee To reduce the number of simulated galaxies, dependent on the depth of the simulated image, the original EGG code uses either a stellar mass cutoff on simulated galaxies, or a UVJ-diagram based selection criteria designed around optical galaxies. The modification in this paper uses the estimated star formation rate (SFR) to calculate a bolometric infrared luminosity using the same empicially calibrated formulas already used in the EGG code.} EGG uses the SkyMaker \citep{skymaker} code to generate survey images from the mock catalogues. The EGG software was used to create a training data set representing the redshift range $0.1\space\leq\space z\space\leq6$. The EGG code generated a number of co-spatial 20 $\mathrm{deg}^{2}$ images for the 4 bands used in this paper with three {\it Herschel} SPIRE bands and one JCMT SCUBA-2 band. Each set of 4 images was cut into non-overlapping subregions covering $424\times424$ $\mathrm{arcsec}^2$, providing a total of 2373 images to train on. 10$\%$ of the generated images were reserved for use as a test set.

\subsection[Observational Data]{Observational Data}\label{Real Data}

A smaller training sample was derived from the small area of overlap within the COSMOS field between the JCMT SCUBA-2 STUDIES large program, and the {\it Herschel} SPIRE maps. The small area of overlap was divided into 144 overlapping images offset from each other in RA and Dec in steps of 12 arcsec. The 12 arcsec offset steps correspond with the pixel scale of the {\it Herschel} SPIRE $500$\microns images and are therefore the smallest possible increment consistent with the lowest resolution images. Each set of images was then flipped and/or rotated to augment the data, to produce 8 images in total at each shifted position. This procedure resulted in an overall dataset containing 1156 images. The 144 non-flipped, non-rotated original images were removed from the training set to be used as a test set, leaving 1008 images for training and ensuring that the images used for testing differed as much as possible from the training set. 

\section[Network Architecture]{Network Architecture}\label{Network Architecture}

The generator network of the auto-encoder is based on that used by the GalaxyGAN code \citep{GalGAN}. The CNN presented in this paper {\bfreferee differs from} GalaxyGAN, and other previous works in two significant ways: (i) the use of a more computationally expensive loss function, that is better designed to extract the individual features of interest, and (ii) no discriminator network is used. 

The network processes the two training sets independently, in succession. Each epoch\footnote{The term \textit{epoch} refers to a complete pass over the combined observed and simulated training data sets.} begins by training on the entire simulated training set, before training on the observed data set three times in succession. The aim was to have the network learn the key structural features of the sub-mm images on the simulated data before using the observed data to fine-tune the network to handle any small differences between observed and simulated images. Each training set was randomised before each run through the data. Due to simulation differences in the flux distribution the observed data were renormalised before training on them to ensure a comparable flux distribution to the simulations.

\subsection[Generator]{Autoencoder}\label{Autoencoder}

The architecture presented in this paper uses a U-net configuration \citep{UNET}. The outputs from each convolutional layer in the encoder network are concatenated with the inputs of their corresponding layer in the decoder network. This helps to prevent the overall network output from diverging substantially from its input. It takes as its input the three {\it Herschel} SPIRE bands ($250$\microns, $350$\microns and $500$\microns) images and is trained to produce an output image that closely matches a target image consistent with the single JCMT SCUBA-2 $450$\microns band. All activation functions in the CNN are LeakyReLU: 

\begin{equation}
f(x) = 
\begin{cases}
\text{$\alpha x$, $x<0$}\\
\text{$x$, $x \geq 0$} 
\end{cases}
\label{eq:l-relu}
\end{equation}
 except for the final layer where a sigmoid function 
 is used:
\begin{equation}
    f(x) = \frac{1}{1 + e^{-x}}\,.
\label{eq:sigmoid}
\end{equation}

 LeakyReLU was chosen as the activation functions over the ReLU function, as the zero-gradient nature of the ReLU function at $x<0$ can cause "dead neurons" in the network. The sigmoid function in the final layer ensures a well constrained output range with continuous coverage. Batch normalisation is included after each convolutional layer to regularise their outputs, which enhances the overall stability of the network and its predictive performance on unseen input data \citep{BN}. For similar reasons, dropout layers are used to randomly disable training 50\% of the kernel weights in the first three layers of the decoder network \citep{dropout_JMLR:v15:srivastava14a}. 

\begin{figure*}
\includegraphics[scale=0.3]{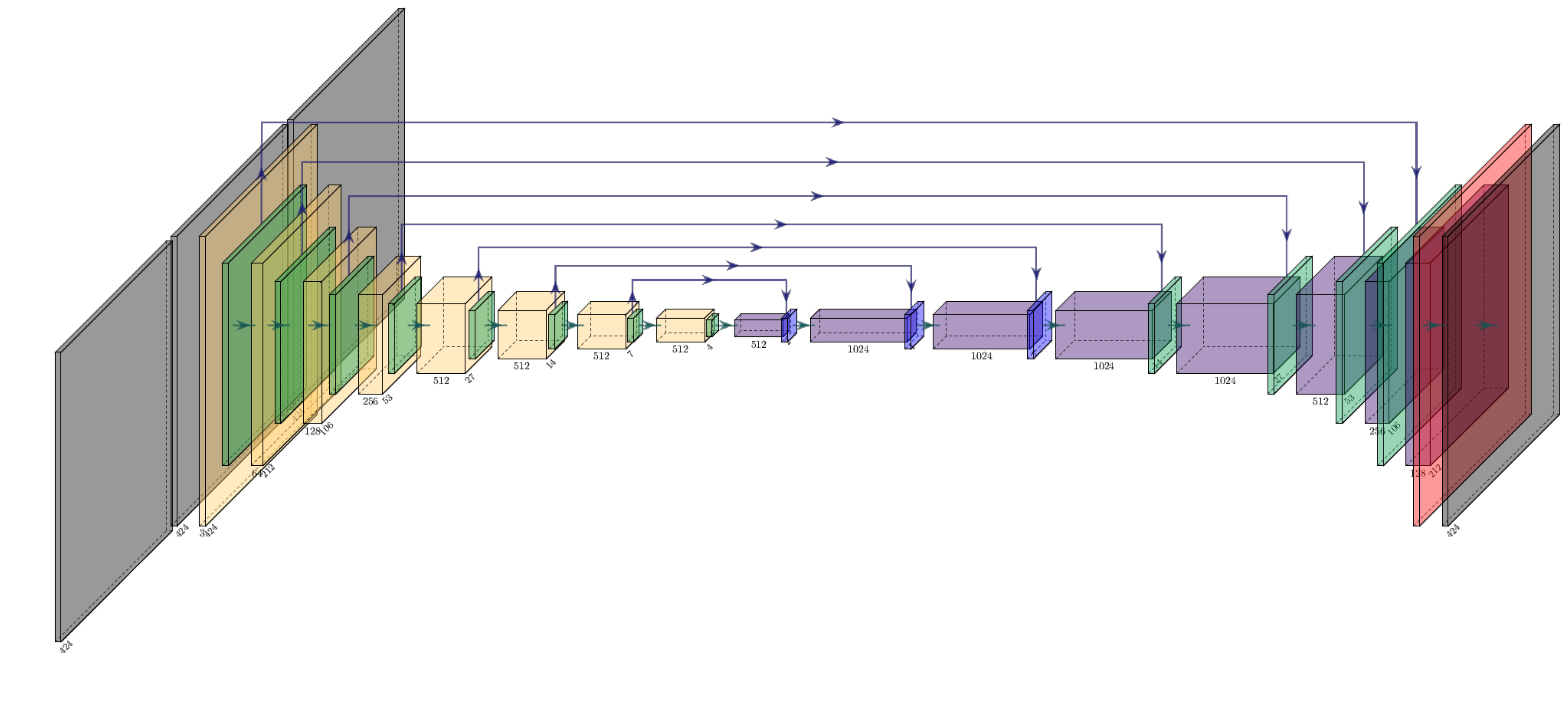}
\caption{{\bfreferee Schematic of the autoencoder used in this work. The yellow boxes represent convolutional layers. The purple boxes represent de-convolutional layers. The green boxes represent combined batch normalisation and LeakyReLU activation function layers. Blue boxes represent a sequence of batch normalisation, dropout and LeakyReLU activation layers. The red box represents the sigmoid activation function in the final layer, and the grey boxes are the input/output images.}}
\label{fig:Network}
\end{figure*}

 The architecture of the CNN is described in table \ref{tab:generator} and a schematic shown in fig. \ref{fig:Network}. Using this architecture requires that the pixel dimensions of the input and ouput images match. However, the {\it Herschel} SPIRE $250$\microns, $350$\microns, and $500$\microns image pixel scales are 6", 8.33", and 12" respectively, while the JCMT SCUBA-2 images have a pixel scale of 1". Accordingly, since the input and output images represent equal areas on the sky, a 2-D linear interpolation routine from the {\sc SciPy} Python package \citep{2020SciPy-NMeth} was used to subsample the input {\it Herschel} images. 

\begin{table*}
\centering
\caption{Autoencoder network architecture. Layers 1-8 comprise the encoder, while layers 9-16 comprise the decoder. The output of the encoder network is a $2\times2\times512$ element tensor embedding of the input image, which is used as the input to the decoder network. The outputs from each convolutional layer in the encoder network are concatenated with the inputs of their corresponding layer in the decoder network. All layers use convolutional kernels $4\times4$ pixel extent in the width and height dimensions. The output of layer 8 encodes the embedding for this auto-encoder network.}
\label{tab:generator}
\begin{tabular}{|l||l|l|l|l|l|l|l|l|}
\hline
	 Layer & Input dimensions & Kernels & Part 1 & Part 2 & Part 3 & Part 4 & Activation & Output dimension\\
	 \hline
	 1 & $424\times 424\times 3$ & $64$ & Conv. & BatchNorm. & & & LeakyReLU & $212\times 212\times 64$\\
	 \hline
	 2 & $212\times 212\times 64$ & $128$ & Conv. & BatchNorm. & & & LeakyReLU & $106\times 106\times 128$\\
	 \hline
	 3 & $106\times 106\times 128$ & $256$ & Conv. & BatchNorm. & & & LeakyReLU & $53\times 53\times 256$\\
	 \hline
	 4 & $53\times 53\times 256$ & $512$ & Conv. & BatchNorm. & & & LeakyReLU & $27\times 27\times 512$\\
	 \hline
	 5 & $27\times 27\times 512$ & $512$ & Conv. & BatchNorm. & & & LeakyReLU & $14\times 14\times 512$\\
	 \hline
	 6 & $14\times 14\times 512$ & $512$ & Conv. & BatchNorm. & & & LeakyReLU & $7\times 7\times 512$\\
	 \hline
	 7 & $7\times 7\times 512$ & $512$ & Conv. & BatchNorm. & & & LeakyReLU & $4\times 4\times 512$\\
	 \hline
	 8 & $4\times 4\times 512$ & $512$ & Conv. & BatchNorm. & & & LeakyReLU & $2\times 2\times 512$\\
	 \hline
	 
	 9 & $2\times 2\times 512$ & $512$ & De-Conv. & BatchNorm. & Dropout & Concat 7 & LeakyReLU & $4\times 4\times 1024$\\
	 \hline
	 10 & $4\times 4\times 1024$ & $512$ & De-Conv. & BatchNorm. & Dropout & Concat 6 & LeakyReLU & $7\times 7\times 1024$\\
	 \hline
	 11 & $7\times 7\times 1024$ & $512$ & De-Conv. & BatchNorm. & Dropout & Concat 5 & LeakyReLU & $14\times 14\times 1024$\\
	 \hline
	 12 & $14\times 14\times 1024$ & $512$ & De-Conv. & BatchNorm. & & Concat 4 & LeakyReLU & $27\times 27\times 1024$\\
	 \hline
	 13 & $27\times 27\times 1024$ & $256$ & De-Conv. & BatchNorm. & & Concat 3 & LeakyReLU & $53\times 53\times 512$\\
	 \hline
	 14 & $53\times 53\times 512$ & $128$ & De-Conv. & BatchNorm. & & Concat 2 & LeakyReLU & $106\times 106\times 256$\\
	 \hline
	 15 & $106\times 106\times 256$ & $64$ & De-Conv. & BatchNorm. & & Concat 1 & LeakyReLU & $256\times 256\times 128$\\
	 \hline
	 16 & $256\times 256\times 128$ & $1$ & De-Conv. & & & & Sigmoid & $424\times 424\times 1$\\
	 \hline

\end{tabular}
\end{table*}

\subsection[Loss function]{Loss Function}\label{Loss Function}

The common approach for deconvolution when designing the loss function for GANs \citep[e.g.][]{GalGAN, Moriwaki_2021} is to combine the loss $L_{\mathrm{disc}}$ from a  discriminator network with a simple $L_{1-\mathrm{loss}}$ between the encoder-decoder network output $y_{\mathrm{predicted}}$ and a target image $y_{\mathrm{true}}$.

\begin{align}
    L &= L_{1-{\mathrm{loss}}}+ L_{\mathrm{disc}} \notag \\
    &=\sum{|y_{\mathrm{true}} - y_{\mathrm{predicted}}|} + L_{\mathrm{disc}}
    \label{eq:L1-loss}
\end{align}

 The approach in this paper is different. While the CNN presented retains the $L_{1-\mathrm{loss}}$ as part of the overall loss function, it is not the main component. The main goal of the CNN in this paper is to super-resolve the sub-mm telescope PSF. To achieve this, a novel, custom loss function was designed to better target the data features of interest. In particular, this multifaceted loss function focuses on the differences between the fluxes of any point sources that are identified in corresponding pairs of generated and target images. 

 The loss computation uses the {\sc Photutils} Python package \citep{photutils} to identify point sources within the target or generated images and extract fluxes from circular apertures with 10 arcsec radius, centred on the identified source locations.
 
 The loss is computed by comparing the fluxes extracted from the generated and target images, but the details of the computation are different when training on the simulated and observed training sets.

\subsubsection{Training on simulated data}
When the network is training on simulated data, the locations of sources with {\bfreferee signal-to-noise ratios of} S/N $> 3$ are derived using the target image only. Fluxes are extracted from both the target and generated images using apertures corresponding to the target image locations. 

 The loss is computed as the sum over all apertures of the absolute difference between the extracted flux in the target and generated image.

\begin{equation}
    L_{\mathrm{train}}^{\mathrm{sim}} = \displaystyle\sum_{i=1}^{N_{s}^{\mathrm{target}}}|f_{i}^{\mathrm{target}} - f_{i}^{\mathrm{generated}}|
\end{equation}
where $N_{s}^{\mathrm{target}}$ is the number of point sources that are identified in the target image, $f_{i}^{\mathrm{target}}$ is the flux extracted from the $i$th aperture in the target image and $f_{i}^{\mathrm{generated}}$ is the flux extracted from the $i$th aperture in the generated image.

\subsubsection{Training on observed data}

When the network is training on observed data, the locations of source are derived for both the target and generated images. For the target image the source identification criterion remains S/N $> 3$, but for the generated image, this threshold is relaxed and all sources with S/N $> 1$ are identified. {\bfreferee The disparity in S/N detection limits used originates in the different purposes of the loss function in the two cases. When detecting sources in the real data the purpose is to replicate the aperture flux in real galaxies. This necessitates that a reasonable lower limit has to be set on sources that are attempted reconstructed. The opposite holds true when detecting sources in the generated data. In this case it is just important to identify spurious flux anomalies that does not correspond to actual sources, necessitating a lower S/N threshold. } Four sets of fluxes are then extracted from both the target and generated images using both sets of apertures. Fluxes are extracted from the target images using apertures from the target and generated images, and vice-versa. 

The loss is computed as
\begin{align}
    L_{\mathrm{train}}^{\mathrm{observed}} =& \frac{1}{N_{s}^{\mathrm{target}}}\displaystyle\sum_{i=1}^{N_{s}^{\mathrm{target}}}|f_{i}^{\mathrm{target}} - f_{i}^{\mathrm{generated}}| \notag\\
    &+ \space\frac{1}{N_{s}^{\mathrm{generated}}}\displaystyle\sum_{j=1}^{N_{s}^{\mathrm{generated}}}|f_{j}^{\mathrm{target} - f_{j}^{\mathrm{generated}}}|
\end{align}
where $N_{s}^{\mathrm{generated}}$ is the number of point sources that are identified in the generated image. The second term explicitly penalises spurious features that appear in the generated image.

\subsubsection{Common loss components}\label{Common loss components}

In addition to those based on the aperture flux differences, three common components also contribute to training loss functions for both the observational and simulated training datasets. The first is the reduced mean of the absolute per-pixel difference between the target and generated images\footnote{This is often referred to as the L1 loss.}.

\begin{equation}
    L_{\mathrm{train}}^{L1} =\frac{1}{N_{\mathrm{pix}}} \displaystyle\sum_{i=1}^{N_{\mathrm{pix}}}|f_{i}^{\mathrm{target}} - f_{i}^{\mathrm{generated}}|
\end{equation}
where $N_{\mathrm{pix}}$ is the number of pixels in either of the images. This component ensures that the loss includes some influence from the bulk of image pixels outside extracted apertures. 

 The second common loss component is the absolute difference between the mean pixel fluxes of the generated and target images. This loss component is designed to encourage the generated image to have an integrated flux similar to that of the target image.

 Finally, the loss includes the absolute difference between the median pixel fluxes of the generated and target images. This component is intended to produce generated images that have a similar distribution of pixel intensities to the target images. Since the majority of image pixels are noise or background dominated, this tends to result in generated images with similar background properties to the targets.

\section[Results]{Results}\label{Results}

\begin{figure*}
\includegraphics[scale=0.58]{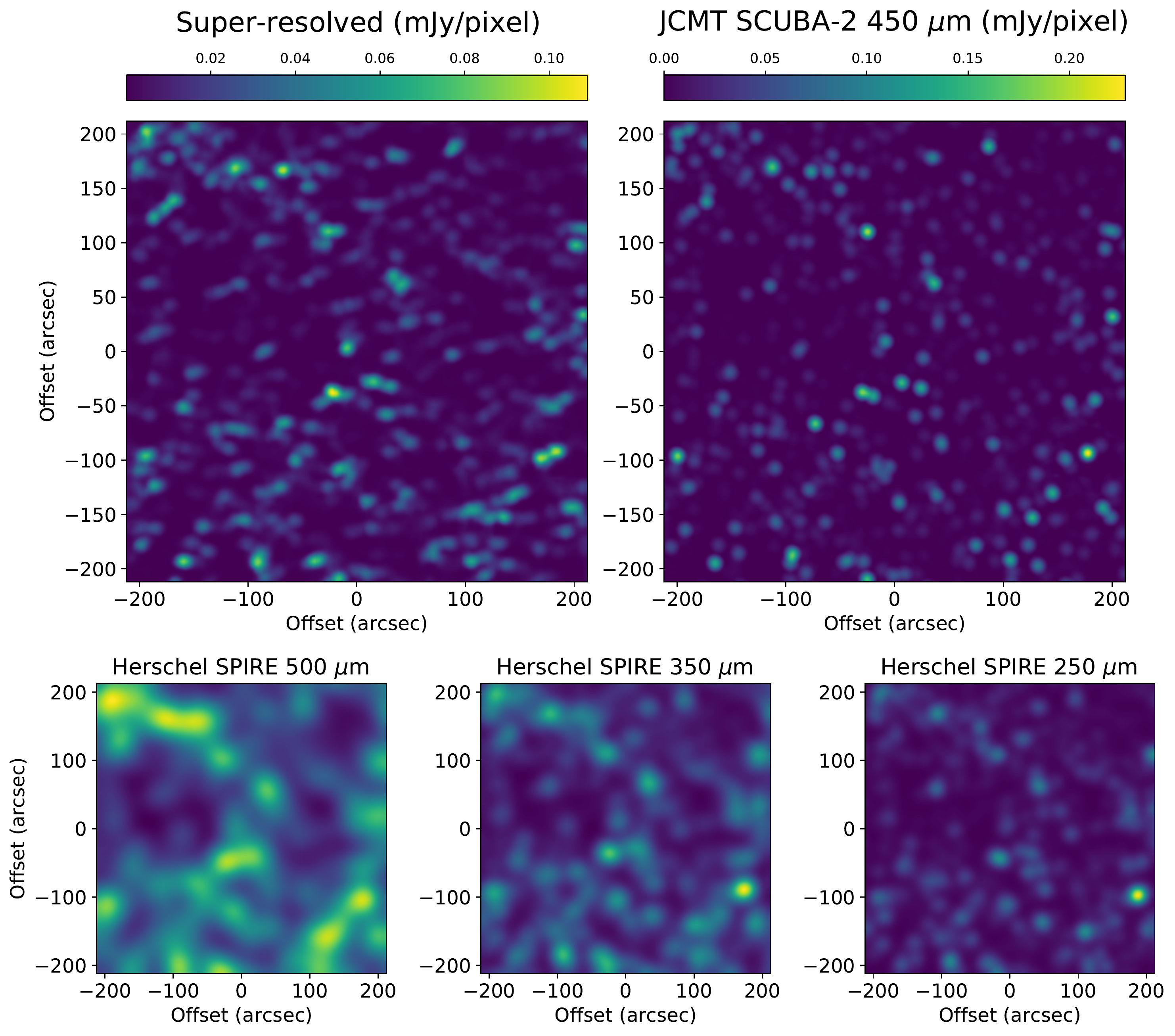}
\caption{The performance of the CNN on simulated COSMOS data from {\it Herschel} SPIRE. Top left is the deconvolved image, top right is the actual JCMT SCUBA-2 $450$\microns image, and the bottom row are the {\it Herschel} SPIRE images. The {\it Herschel} SPIRE images shown here are processed identically to the network inputs. They are 2-D linearly interpolated, and linearly normalised to have pixel values between zero and unity. The simulated JCMT $450$\microns image demonstrates the depth that originates in the high S/N possible with simulations in the lack of discernible noise.}
\label{fig:fake_test}
\end{figure*}

\begin{figure*}
\includegraphics[scale=0.58]{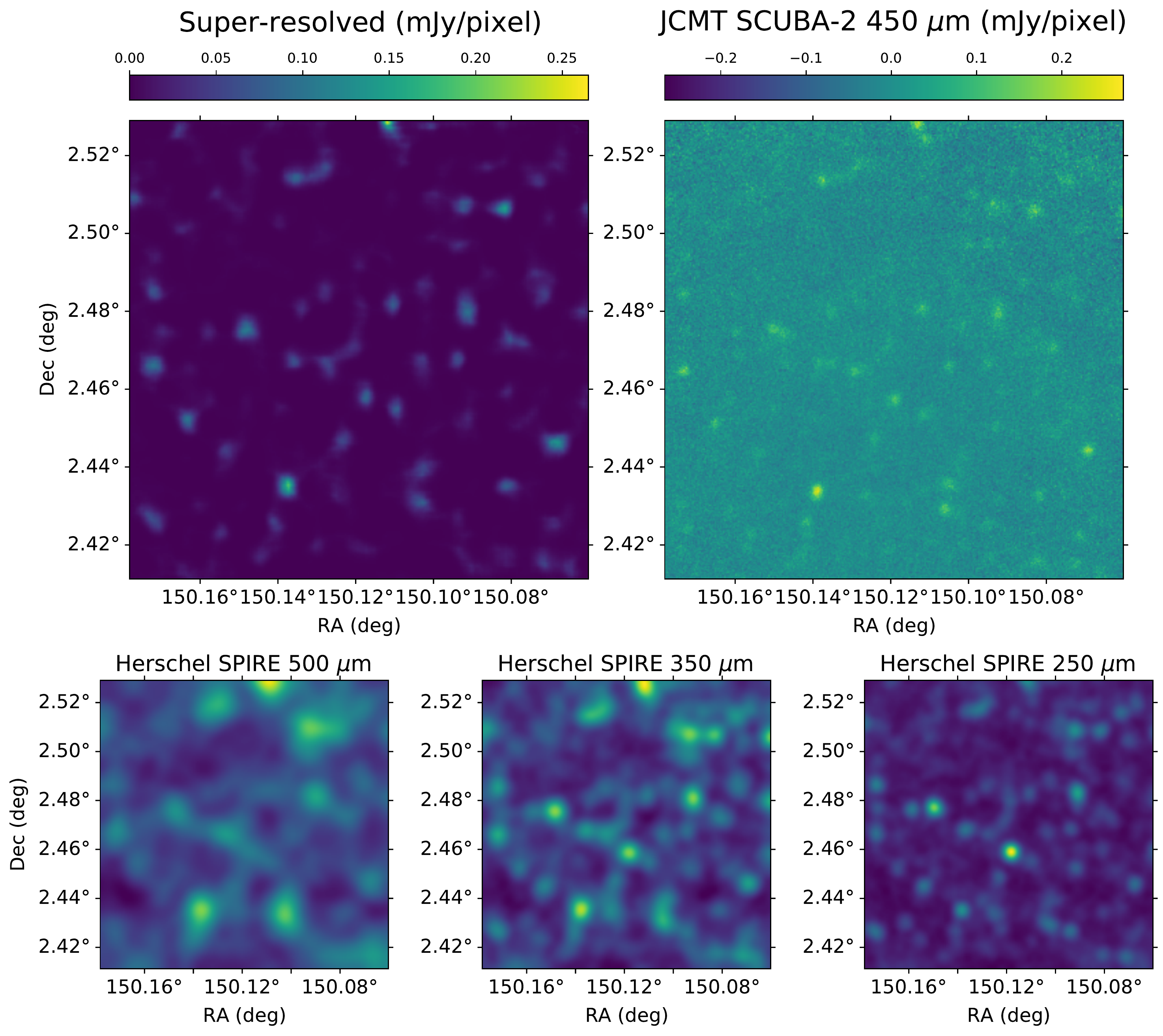}
\caption{The performance of the CNN on real COSMOS data from {\it Herschel} SPIRE. Top left is the deconvolved image, top right is the actual JCMT SCUBA-2 $450$\microns image, and the bottom row are the {\it Herschel} SPIRE images. The {\it Herschel} SPIRE images shown here are processed identically to the network inputs. They are 2-D linearly interpolated, and linearly normalised to have pixel values between zero and unity. The JCMT $450$\microns image shows the noise inherent in real observations, while the super-resolved image shows the power of an autoencoder in reconstructing the JCMT $450$\microns image without the clear noise contribution.}
\label{fig:real_test}
\end{figure*}

\begin{figure*}
\centering
\includegraphics[width=.47\linewidth]{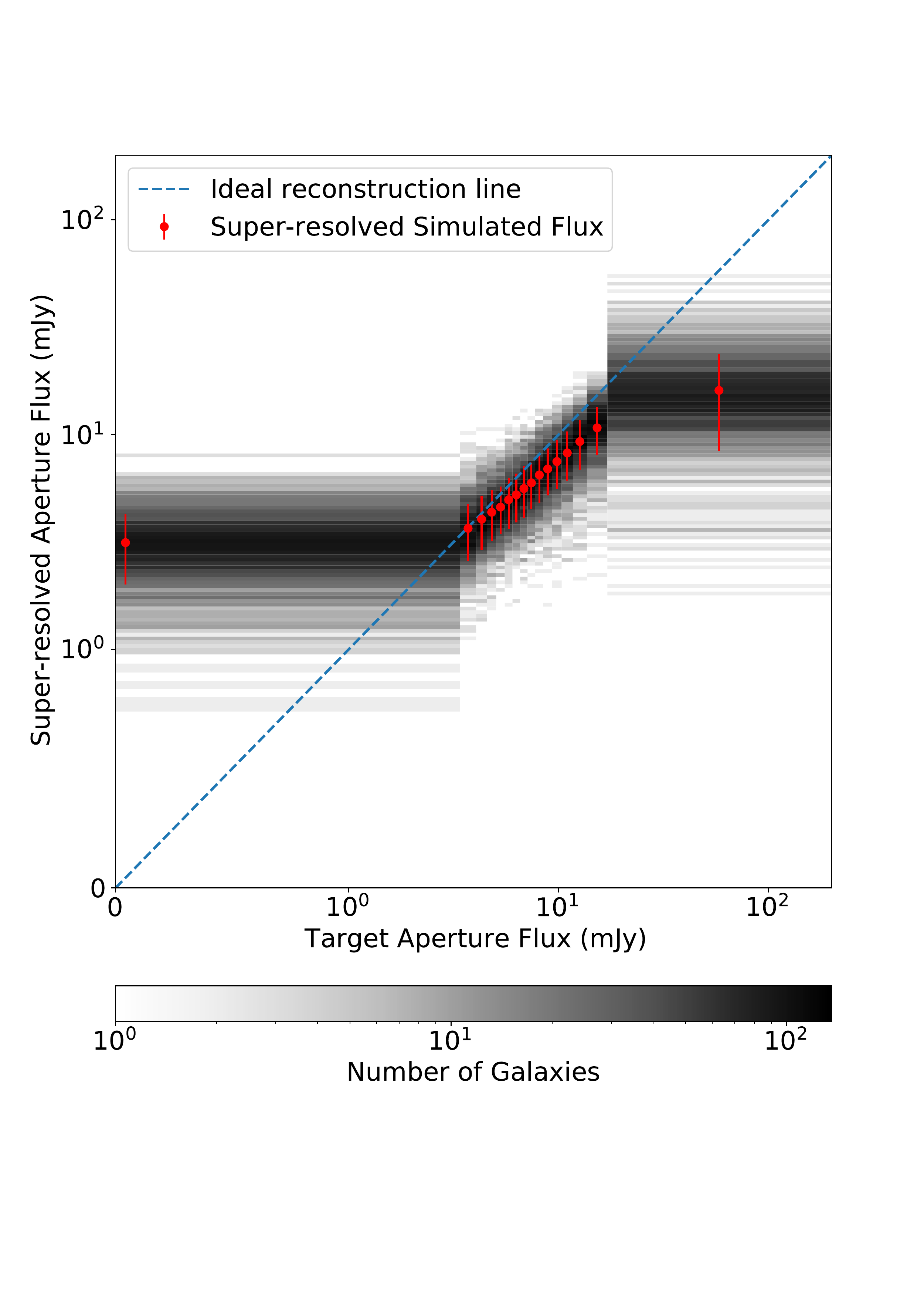}
\includegraphics[width=.47\linewidth]{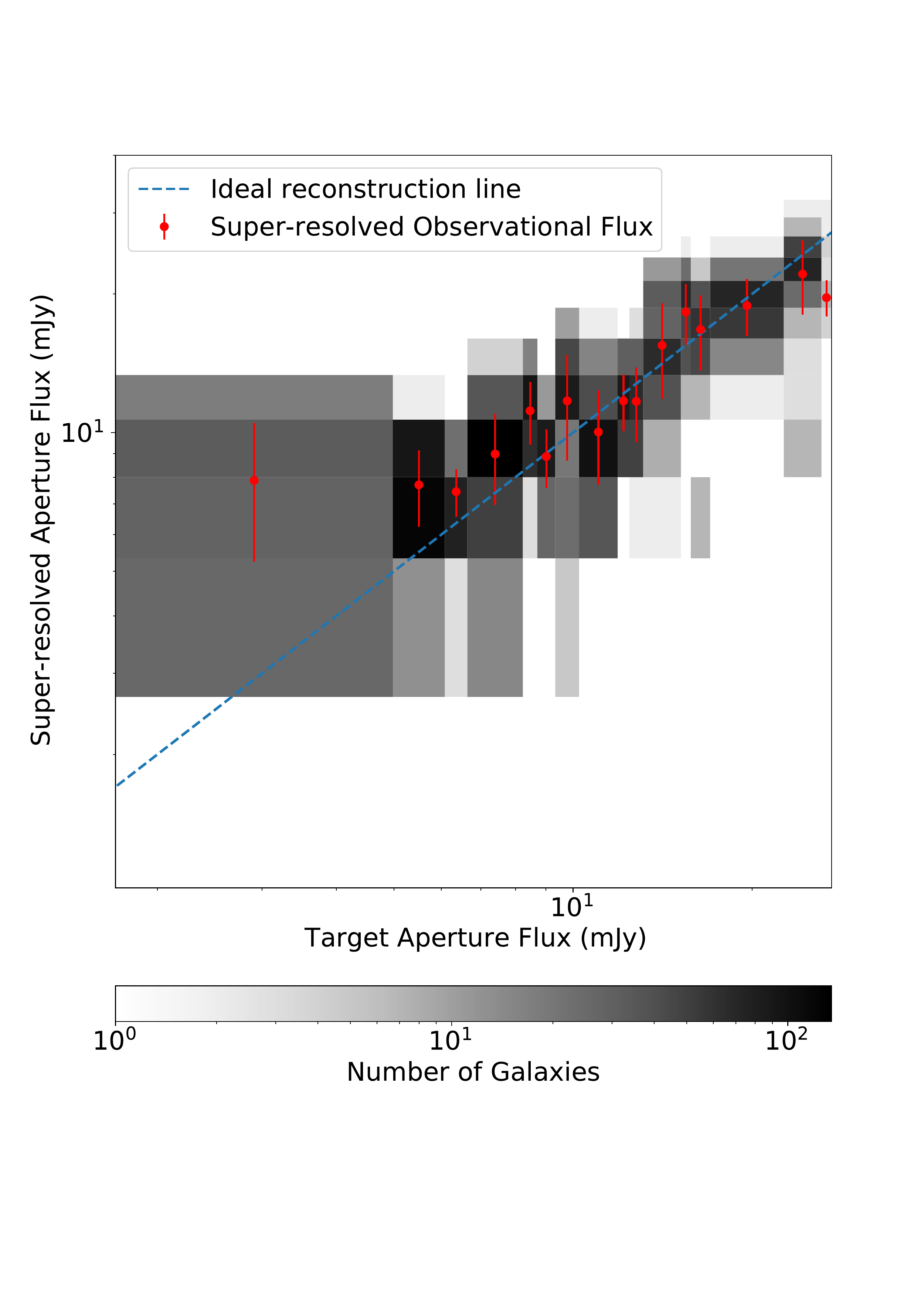}
\vspace*{-2cm}
\caption{\textit{Left panel:} Comparison between fluxes of point sources that are detected in super-resolved simulated images and fluxes extracted from spatially coincident point sources in simulated high resolution target images. Only sources with fluxes exceeding 10 times the background RMS in the super-resolved image are considered. Note that the distribution of background noise is highly non-Gaussian and the significance of any point sources in the super-resolved image, relative to the background level cannot be interpreted in a standard Gaussian framework. The flux is calculated using aperture photometry within a circular aperture of 10" centered on the source locations. The grey histogram shows the number of sub-mm galaxies detected in each bin of target versus super-resolved flux space. The red points and errors show the mean super-resolved aperture flux within each target flux bin and its associated standard deviation, respectively. {\bfreferee The bins are defined to ensure equal numbers of galaxies in each bin, which results in the faintest and brightest bins covering a large logarithmic range. The red data points are located at the bin centres in logarithmic flux space. For clarity, the axes on the left panel are linear below 1\,mJy and logarithmic above this value.} \textit{Right panel:} Same as left-hand panel, but comparing \textit{observed} high resolution JCMT SCUBA-2 $450$\microns images with super-resolved \textit{observed} {\it Herschel} SPIRE counterparts. Note the Eddington bias in the faint fluxes, caused by pre-selection on faint features in the reconstruction (see text).}
\label{fig:real_flux}
\end{figure*}

The CNN presented here was trained and tested using both a pure simulation data set and on a data set combining simulated and observed data. In Fig. \ref{fig:fake_test} the performance on a pure simulation data set is demonstrated. The performance of the network on the combined simulated and observed data can be seen in Fig. \ref{fig:real_test}. It is clear that the target images for the simulated data contain more discernible sources than the real JCMT images do. This is likely due to the reduced noise in the simulated data. Figs. \ref{fig:fake_test} and  \ref{fig:real_test} show the {\it Herschel} bands, as they are provided to the network, post-normalisation. The network has no effective way of recreating the noise inherent in real observations. The median, and $L_{1-\mathrm{loss}}$ components of the loss function, should drive the {\bfreferee network} to represent the mean noise level as a quasi-uniform background flux, the spatially varying nature of the real data noise will not be reproduced. The noise in the real background-subtracted JCMT {\bfreferee image} is distributed around zero. This drives the background level generated by the CNN to be very close to zero, but it can never be negative because the sigmoid activation of the output layer does not allow negative values. The enforced non-negativity of the super-resolved image pixel values also means that the distribution of background noise is highly non-Gaussian and the significance of any point sources in the super-resolved image, relative to the background level cannot be interpreted in a standard Gaussian framework.  
The right-hand panel of Fig.\,\ref{fig:real_flux} shows an Eddington-like bias in the reconstructed fluxes at the faint end \citep{Eddington1913}, caused by pre-selecting faint features in the reconstruction. The matching with the JCMT observations uses a lower threshold for features.

\begin{figure*}
\includegraphics[scale=0.6]{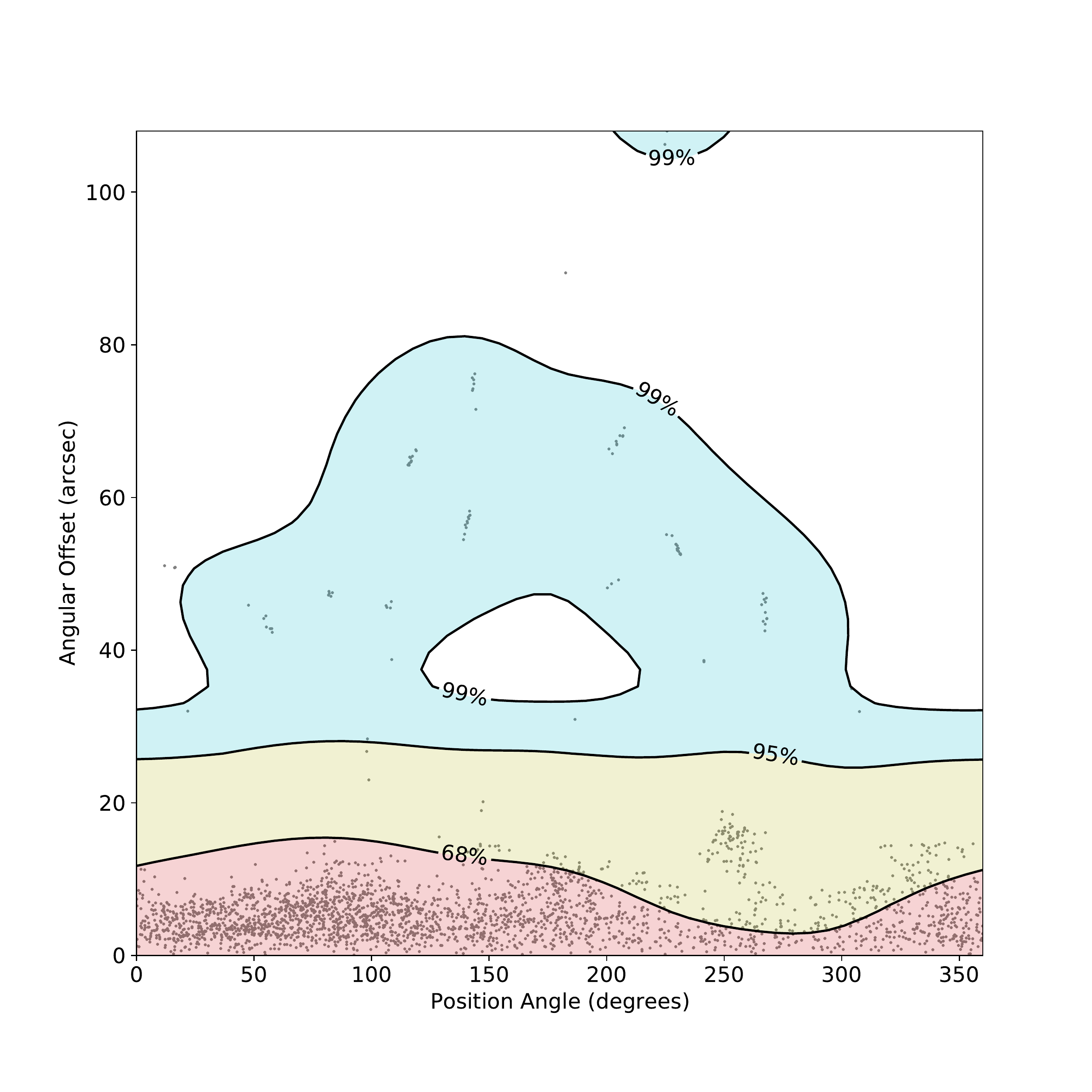}
\caption{Angular distance versus position angle for offsets of extracted sources identified on the super-resolved images and the nearest counterpart point source in coincident SCUBA-2 imaging. Position angles are defined in degrees anti-clockwise from West. Grey markers correspond with individual objects that are identified in the super-resolved images. Contours showing the 68th, 95th and 99th percentiles are overlaid. Overall, the astrometric accuracy is excellent with 99\% of the super-resolved objects having offsets $\lesssim15$ arcseconds. There is some evidence for clustering of offsets to the Northwest.}
\label{fig:astro_dir}
\end{figure*}

 Fig. \ref{fig:astro_dir} shows the astrometric error on the predicted locations of all of extracted sources detected in the super-resolved image. {\bfreferee The natural intuition from single-dish observations is that the positional uncertainty of a point source should be approximately $0.6\theta_{\mathrm{FWHM}}/$(S/N) where $\theta_{\mathrm{FWHM}}$ is the beam full-width half maximum and S/N is the signal-to-noise ratio of that source \citep[e.g.][]{Ivison2007}. However, in this case, the reconstructed map is not a single-dish observation, even though it resembles one. The astrometric uncertainty is a non-trivial product of the map reconstruction, and therefore it is something that must be determined directly from the comparison with the truth data.}
 {\bfreferee A further complicating factor is that the pixel scales of the originally used {\it Herschel} SPIRE images are 6", 8.33" and 12", and that of the JCMT images is 1", while the network uses images of the size $424\times424$ pixels. As $424$ is not divisible by either of the {\it Herschel} SPIRE pixel scales, this will cause minor differences in the exact astrometric alignment of the images fed to the network causing small astrometric errors}.  Furthermore, the misalignment of the pixels in the {\it Herschel} SPIRE data, due to the individual pixel scales not being integer multiples of one another might cause additional issues. However, {\bfreferee we find that} the astrometric accuracy is {\bfreferee often} better than the pixel scale of the {\bfreferee {\it Herschel} SPIRE 500\microns images, which is the closest equivalent image to the JCMT SCUBA-2 images}.

 Observed flux is a key characteristic of observed galaxies. A significant portion of the loss function was designed to target the recovery of this observable. Fig. \ref{fig:real_flux} shows the relationship between the super-resolved fluxes and the target fluxes of all sources brighter than 10 times the background flux RMS in the super-resolved image, for all the test image pairs. To compute an absolute flux calibration for the super-resolved point-source fluxes, the JCMT SCUBA-2 $450$\microns and super-resolved images are both convolved with a 2D Gaussian with a FWHM of 36.3 arcsec. A mask is generated to isolate the brightest pixels in the {\it Herschel} $500$\microns image and the pixel fluxes at the unmasked locations are compared with corresponding pixel fluxes in the two convolved images. Two linear scaling relations are found which map the pixel fluxes in the {\it Herschel} $500$\microns image to those in the convolved JCMT SCUBA-2 $450$\microns and super-resolved images. Finally, a direct calibration from the super-resolved image flux to the corresponding high-resolution is derived by concatenating these two linear mappings.
 
\begin{figure*}
\includegraphics[scale=0.6]{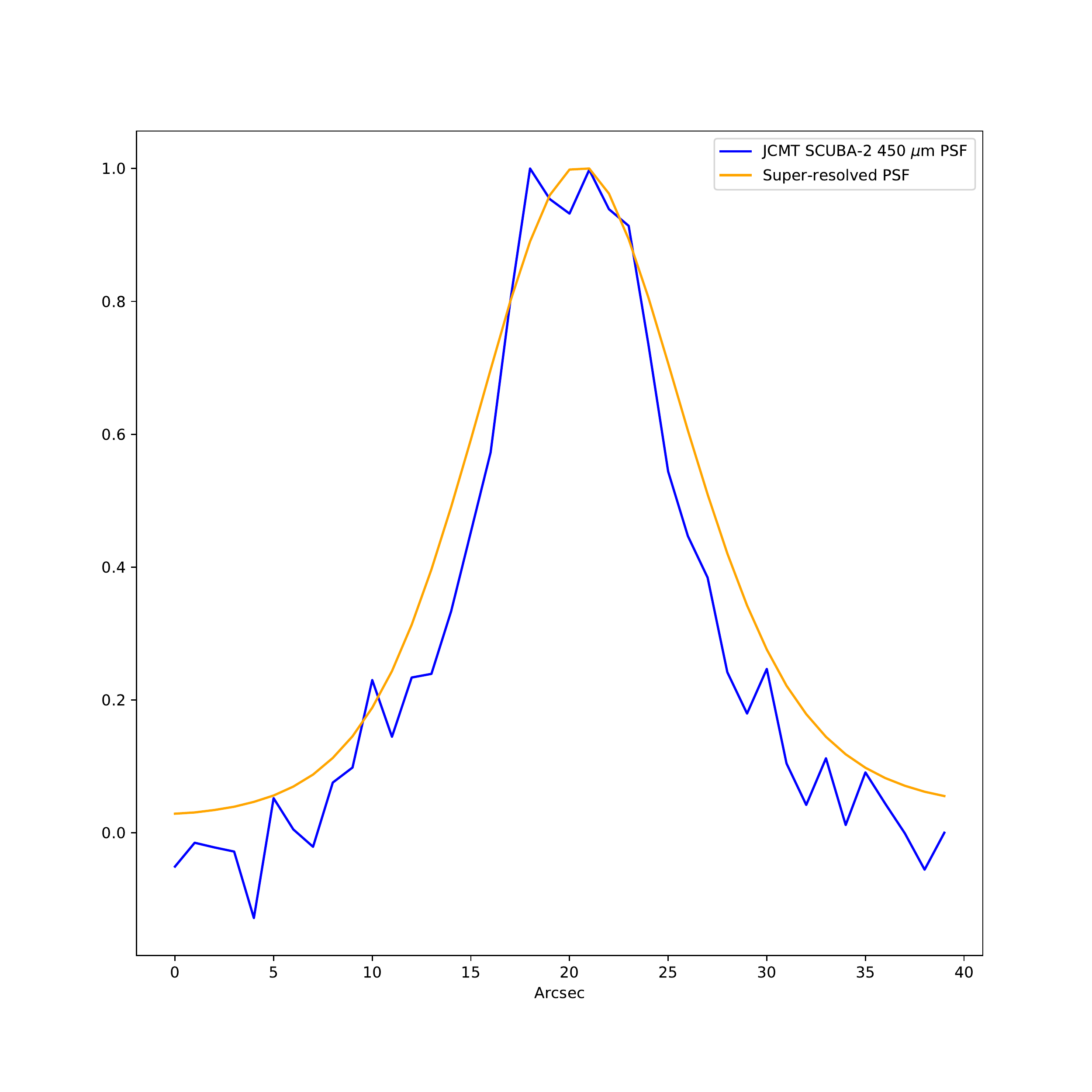}
\caption{{\bfreferee The stacked PSF of the real JCMT SCUBA-2 $450$\microns sources (blue), and the super-resolved sources (orange). The fluctuations in the real sources originate from the small sample of real images that can be used for training and validation. This causes a large overlap of sky area repeating the same source multiple times. The smoothness of the super-resolved PSF is achieved by the final activation layer suppressing the noise in the generated images.}}
\label{fig:psf_profile}
\end{figure*}

 While the network does seem to slightly underestimate the calibrated flux for the simulated sources, the results for observational data show good promise. It is worth noting that due to the substantial overlap between the sky areas covered by the individual test images, many of the extracted fluxes correspond to the the same sub-mm galaxy, seen in a different image. Thus the source distribution might not be entirely representative. For bright sources identified in both the simulated and observational datasets, an approximately 1:1 correlation between the calibrated, super-resolved fluxes and their counterparts in the target images is evident, albeit with some scatter. This correlation implies that the fluxes of bright sources can be reliably extracted from super-resolved images. Note that the custom loss function is designed to recover the total flux within a 10" aperture{\bfreferee . The pull is defined as $g=|x-\mu|/\sigma$, with $x$ being the expected value, $\mu$ being the mean value of the bin, and $\sigma$ being the standard deviation. The pull has been calculated for the reconstructed source fluxes in table \ref{tab:pull}, where it is shown that the pull for sources between 9 and 24 mJy with only one exception varies between 0.11 and 0.65. A stacking of the reconstructed sources reveals that the reconstruction has a PSF profile very similar to that of the target data (see fig. \ref{fig:psf_profile}). This is achieved with} only the $L_{1-\mathrm{loss}}$ part of the loss function {\bfreferee trying} to replicate the PSF shape.
 
\begin{table}
\centering
\caption{{\bfreferee The pull calculated for the reconstructed sources shown in the righthand panel of fig. \ref{fig:real_flux}.}}
\label{tab:pull}
\begin{tabular}{|l|l|}
\hline
	 Flux (mJy) & Pull \\
	 \hline
	 2.91 & 1.89 \\
	 \hline
	 5.5 & 1.51 \\
	 \hline
	 6.36 & 1.22 \\
	 \hline
	 7.39 & 0.79 \\
	 \hline
	 8.47 & 1.55 \\
	 \hline
	 9.01 & 0.11 \\
	 \hline
	 9.77 & 0.64 \\
	 \hline
	 11.03 & 0.43 \\
	 \hline
	 12.15 & 0.27 \\
	 \hline
	 12.77 & 0.50 \\
	 \hline
	 14.11 & 0.37 \\
	 \hline
	 15.47 & 1.02 \\
	 \hline
	 16.38 & 0.12 \\
	 \hline
	 19.61 & 0.55 \\
	 \hline
	 24.32 & 0.55 \\
	 \hline
	 26.67 & 3.99 \\
	 \hline

\end{tabular}
\end{table}
 
\begin{figure*}
\centering
\includegraphics[width=.47\linewidth]{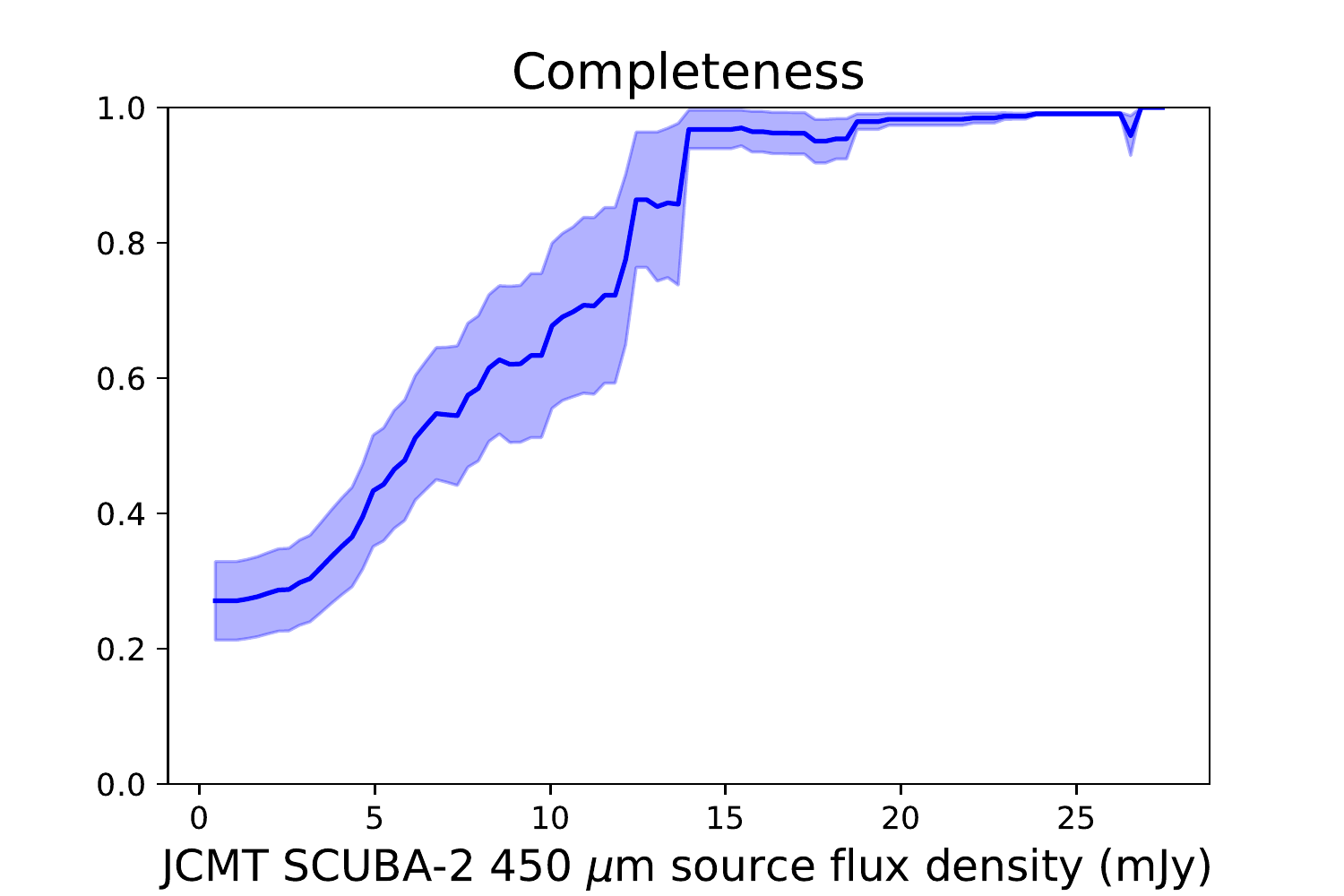}
\includegraphics[width=.47\linewidth]{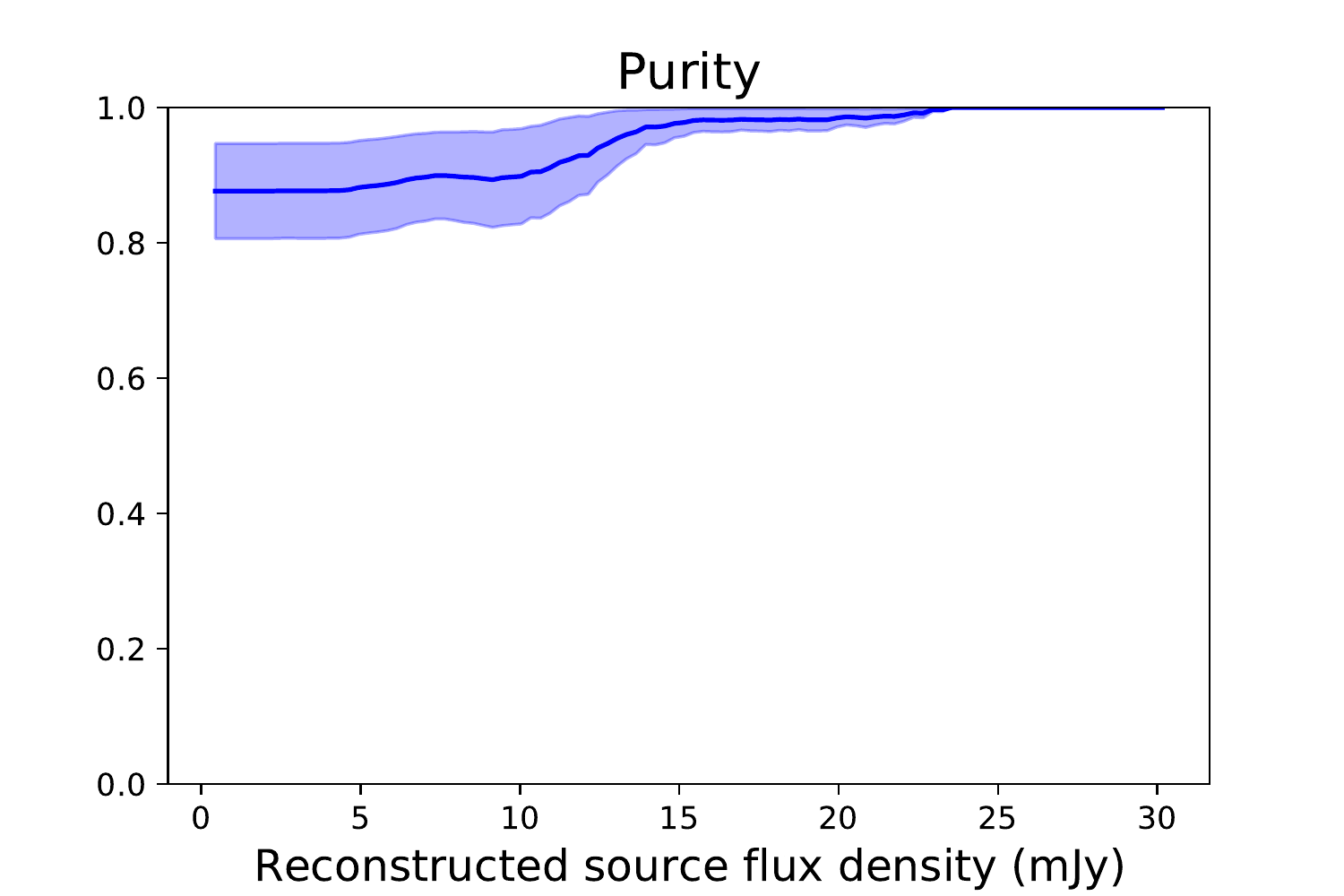}
\caption{{\bfreferee \textit{Left panel:} Completeness of recreated sources inside 10" of detected SNR $\geq5$ sources in the JCMT SCUBA-2 450\microns STUDIES survey maps. \textit{Right panel:} Purity of sources detected in the reconstructed image as compared to sources with SNR $\geq5$ in the JCMT SCUBA-2 450\microns STUDIES survey. In both cases, the horizontal axis shows the minimum flux density threshold under consideration. The shaded areas show the extent of the $\pm1\sigma$ binomial uncertainties.}}
\label{fig:Purity}
\end{figure*}

{\bfreferee The completeness (also known as recall) and purity (also known as reliability) of the reconstructed sources are shown in fig. \ref{fig:Purity}. Completeness is defined as $TP/(TP+FN)$ where $TP$ is the number of true positives and $FN$ is the number of false negatives; purity is $TP/(TP+FP)$ where $FP$ is the number of false positives. Completeness is evaluated considering a set of ``real'' sources with SNR $\geq5$ in the JCMT SCUBA-2 450\microns STUDIES survey maps. Sources that are detected in the generated maps are considered to be true positives if they fall within 10" of a real source and false negatives otherwise. On the other hand we evaluate purity by considering the set of all ``potential'' sources that are detected in the generated maps. Potential sources that fall within 10" of a real source are deemed to be true positives and all other potential sources are counted as false positives.
The completeness is $>$ 95\% at sources brighter than 15\,mJy, and above 60\% at 10\,mJy. The purity does not drop below 87\% at any point. Note that our reconstruction is remarkably complete even below the formal 500\microns blank-field confusion limit for {\it Herschel} SPIRE (table \ref{tab:telescope_instruments}).}
 
\section[Discussion]{Discussion}\label{Discussion}

While many comparable neural networks  \citep[e.g.][]{GalGAN, image_restoration_cycle,Moriwaki_2021} use the output of a discriminator network as part of their loss functions this paper adopts a different approach.  Recall that the training objective of a discriminator component in a GAN is to effectively distinguish between images that have  been artificially generated or processed and images that are genuine or pristine. However, in order to make this distinction, it may rely on features of the images that a human interpreter might consider unimportant. In this paper, the most important objective from a human perspective is for the neural network to recover the locations and fluxes of the genuine point sources in the target data. However, from the perspective of a CNN it may be that the data sets used in this paper (see Figs. \ref{fig:fake_test} and \ref{fig:real_test}) differ most significantly in their noise characteristics. It is therefore possible that a discriminator network would realise its training objective more effectively by focusing its attention on the fine details of the image noise, and disregarding the point source properties like astrometry and flux. Conversely, by using a hand engineered loss function, the network presented in this paper can be forced to focus on the image features that are most critical for the overall objective of super-resolving low resolution images.

Fig. \ref{fig:astro_dir} plots the offset distances and position angles between the locations of identifiable point sources in the super-resolved {\it Herschel} SPIRE data and the nearest sub-mm galaxy location in the corresponding JCMT SCUBA-2 imaging. Overall reconstruction accuracy is excellent, with {\bfreferee a purity calculated at above 87\% at all reconstructed source flux densities, and completeness above 95\% at target source flux densities above 15 mJy}. Nonetheless, some small offsets between the reconstructed and target source positions are apparent. These offsets are likely caused by the different pixel scales for the different {\it Herschel} SPIRE bands, and the JCMT data. These pixel scales are not exact multiples of each other and so pixels from the different image bands intercept flux from different parts of the sky, and may encode information about different subsets of the true source distribution. Even after interpolation, the sources which fall close to the edge of a pixel in the lower resolution bands have inherently uncertain positions, which is likely reflected in the CNN output. Further, the uncertain alignment of in particular the $350$\microns band might cause problems. The 12" and 6" of the $500$\microns and $250$\microns bands divides into each other, while the 8.33" of the $350$\microns band might cause some uncertainty in source location when the images are shifted during data augmentation. Finally, the redder sources might have higher astrometric uncertainty as they are less represented in the higher resolution {\it Herschel} bands. While further work might reduce this astrometric offset, Fig. \ref{fig:astro_dir} shows a tendency of astrometric precision better than the {\bfreferee 12" pixel scale of the $500$\microns {\it Herschel} SPIRE band}. 

Following this successful proof of concept, there are several obvious next steps. These go beyond this initial analysis, and at least some of these will be presented in future papers.

Firstly, this deconvolution algorithm will be applied to all the {\it Herschel} SPIRE extragalactic survey data sets. For deeper fields with richer multi-wavelength complementary data, the deconvolution can be compared to other approaches that use this supplementary data as a prior \citep[e.g.][]{Hurley2017, Serjeant2019}. 

Secondly, there are enhancements that can be made to the simulations, such as incorporating Galactic cirrus. Furthermore, \cite{Dunne2020} find that foreground large-scale structure can statistically magnify the background sub-mm source counts, so one improvement to the simulations would be to incorporate optical/near-infrared imaging and the effects of weak lensing. The deconvolutions would then be able to make use of the three SPIRE bands and the optical/near-infrared data. In the present analysis, the statistical clustering properties of sub-mm galaxies are implicitly (and non-trivially) used to reconstruct the missing Fourier modes on scales smaller than the point spread function (section \ref{Introduction}), so simulating a wider range of clustered multi-wavelength training data should improve the deconvolution. Strong gravitational lensing could also be included \citep[e.g.][]{Negrello2010}, in which case the network could also encode multi-wavelength information, such as the presence of a foreground elliptical or cluster to signpost possible strong lensing. Extending the simulations and neural net training to a wider multi-wavelength domain has the potential in principle to implicitly incorporate more information than explicit multi-wavelength priors, albeit at a cost of less direct interpretability.

Thirdly, the loss function can be tailored to suit particular science goals. The present analysis represents a particular balance between source completeness, source reliability, flux reproducibility and astrometric accuracy, but other choices are possible. There is no reason to suppose that a single "best" deconvolution to suit all purposes is possible even in principle. Indeed, the balances between angular resolution, point-source sensitivity and large-scale features are usually explicit and deliberate choices in astronomical image processing, driven in each case by the particular science goals \citep[e.g.][]{Briggs1995, Serjeant2003, Smith2019, Dragonfly}. One could imagine optimising the loss function not just for completeness or reliability or some balance thereof, but instead to reproduce the sub-mm galaxy source counts, or make the best estimate of the two-point correlation function of sub-mm galaxies, or reliably detect faint ultra-red sub-mm galaxies. 

\section[Conclusion]{Conclusions}\label{Conclusion}

This paper has shown that it is possible to super-resolve {\it Herschel} SPIRE data using CNNs. In this paper an autoencoder was chosen. A new and innovative loss function was engineered to better replicate the image features of interest.

It is possible to reconstruct both astrometry and source flux using this method with some uncertainty. It is expected that the performance on particularly the source flux would improve with a larger, more varied training set of observed data, reducing the need for simulated data in the training phase. More realistic simulated data might also achieve this goal.
 
Ultimately this method will allow for further exploration of the fields observed by {\it Herschel} SPIRE as a complement to the observations carried out with JCMT and similar telescopes.

\section*{Acknowledgements}
{\bfreferee We thank the anonymous referee for many thoughtful and helpful comments that improved this paper.}
The sub-mm observations used in this work include the STUDIES program (program code M16AL006), archival data from the S2CLS program (program code MJLSC01) and the PI program of \citet[][program codes M11BH11A, M12AH11A, and M12BH21A]{Casey2013}.
The James Clerk Maxwell Telescope is operated by the East Asian Observatory on behalf of The National Astronomical Observatory of Japan; Academia Sinica Institute of Astronomy and Astrophysics; the Korea Astronomy and Space Science Institute; Center for Astronomical Mega-Science (as well as the National Key R\&D Program of China with No. 2017YFA0402700). Additional funding support is provided by the Science and Technology Facilities Council of the United Kingdom and participating universities and organizations in the United Kingdom and Canada. Additional funds for the construction of SCUBA-2 were provided by the Canada Foundation for Innovation. 
The authors wish to recognize and acknowledge the very significant cultural role and reverence that the summit of Maunakea has always had within the indigenous Hawaiian community.  We are most fortunate to have the opportunity to conduct observations from this mountain.

This research has made use of data from HerMES project (http://hermes.sussex.ac.uk/). HerMES is a Herschel Key Programme utilising Guaranteed Time from the SPIRE instrument team, ESAC scientists and a mission scientist.
The HerMES data was accessed through the Herschel Database in Marseille (HeDaM - {\tt http://hedam.lam.fr}) operated by CeSAM and hosted by the Laboratoire d'Astrophysique de Marseille.
The OBSIDs of the {\it Herschel} fields used were: 1342195856, 1342195857, 1342195858, 1342195859, 1342195860, 1342195861, 1342195862, 1342195863, 1342222819, 1342222820, 1342222821, 1342222822, 1342222823, 1342222824, 1342222825, 1342222826, 1342222846, 1342222847, 1342222848, 1342222849, 1342222850, 1342222851, 1342222852, 1342222853, 1342222854, 1342222879, 1342222880, 1342222897, 1342222898, 1342222899, 1342222900 and 1342222901.

This research made use of Astropy,\footnote{http://www.astropy.org} a community-developed core Python package for Astronomy \citep{astropy2013,astropy2018}. 
This research made use of Photutils, an Astropy package for
detection and photometry of astronomical sources \citep{photutils}. Data analysis made use of the Python packages Numpy \citep{harris2020array}, Scipy 
\citep{2020SciPy-NMeth} and Pandas \citep{reback2020pandas,mckinney-proc-scipy-2010}
as well as Tensorflow \citep{tensorflow2015-whitepaper} in Keras \citep{chollet2015keras}.
Figures were made with the Python package Matplotlib \citep{Hunter:2007}.

SS and HD were supported in part by ESCAPE - The European Science Cluster of Astronomy \& Particle Physics ESFRI Research Infrastructures, which in turn received funding from the European Union's Horizon 2020 research and innovation programme under Grant Agreement no. 824064. SS and LL thank the Science and Technology Facilities Council for support under grants ST/P000584/1 and ST/T506321/1 respectively.

\section*{Data Availability}

The observational data used for this paper are the {\it Herschel} SPIRE COSMOS data, the maps for which can be downloaded at {\tt https://irsa.ipac.caltech.edu/data/COSMOS/ images/herschel/spire/}, and the JCMT SCUBA-2 STUDIES data.

\bibliographystyle{mnras}
\bibliography{ref.bib}

\bsp	
\label{lastpage}
\end{document}